\documentclass[11pt]{article}
\usepackage{epsfig,amsmath,ctagsplt,subeqnarray,subfigure,tables,float,floatfig,times}
\usepackage[english]{babel}
\usepackage{fancyhdr}

\begin{document}
\pagestyle{empty}

\begin{center}

\bigskip 
\begin {LARGE} \textbf{Study of
 e/$\gamma$ Trigger \bigskip
\\
for the Electron Calibration Stream } \end {LARGE}

\vskip 3. cm

   \begin{small}
M.~Verducci{\small$^{^{\,a}}$}, R.~Hawkings{\small$^{^{\,b}}$}.\\
\vspace{0.5cm}
{\small{$^{a}$}} {\sl{European Organization for Nuclear Research
  (CERN) and CNAF Bologna}},\\
{\small{$^{b}$}}  {\sl{European Organization for Nuclear Research
  (CERN)}}.\\

  \end{small}
 \vskip 3. cm


\end{center}

\begin{center}
 \begin {large} 
 \textbf{Abstract} \end {large} 
\end{center}

This note describes a study of the possibilities for selecting an electron 
calibration stream using the High level electron trigger 
(Level 2 and the event filter).
Using the electromagnetic calorimeter reconstruction and the track
reconstruction algorithms, an evaluation has been performed of the
selection efficiencies and purities for different physics channels
with single or double electrons in the final state.
A study of the calibration stream composition, including the background from 
QCD processes, is presented.

\vspace{1.0cm}
{

 }

\newpage


\pagestyle{plain}
\pagenumbering{arabic}
\setcounter{page}{1}



\section{Introduction}

According to the ATLAS Computing Model, we expect to have four event data 
streams produced from the HLT (high level trigger) system: a physics stream 
dedicated to the reconstruction of the full ATLAS event sample; an express 
line stream to rapidly monitor calibration and reconstruction quality on a 
subset of events before the complete reconstruction is run on the physics 
stream, and to look for interesting and unusual events; a
pathological stream to study  events causing problems for the HLT; and finally 
a calibration stream, processed rapidly and used to obtain the calibration 
constants for the physics stream reconstruction. The calibration stream
will itself be divided into several sub-streams with different types of events.
In particular, this study is devoted to the calibration stream for the
electrons \cite{nota1}.

Using the High Level Trigger algorithms, described in detail in section \ref{soft}, we have studied  the expected rates and purity of the electron stream 
as a function of the luminosity of the LHC.
   
\section{Tools}
In this section all the software tools used are briefly described.
All the selection cuts applied and their meanings are defined.
Moreover, there is a description of the datasets used in the
analysis.

In detail, all the cuts applied for each trigger hypothesis
algorithm are summerized in the table \ref{algo}.
The values of the thresholds are those presented in a recent talk on
the trigger performance \cite{cibran_1}. 

\subsection{Software Tools \label{soft}}
The selection at the High Level Trigger (HLT) \cite{hlt} is seeded by the
information obtained at Level~1 (i.e. Regions-of-Interest, RoI). The 
level~2 trigger reduces drastically the number of events with a relatively 
small latency, using the information contained in the RoI, while the final 
trigger level, the event filter, having fewer constraints on the latency, 
refines the selections using the full event information.
After the HLT, we obtain typically ``Physics Objects''; these are: muons, 
electrons, jets.
In this case we studied the Trigger Menu for the electron objects, in
particular the selection signatures: e25i, selecting single electrons 
with a threshold 
of 25 GeV, and 2e15i, for di-electron events containing two electron 
candidates with at least 15 GeV.
The selection criteria for electrons include a shower-shape analysis in the electromagnetic calorimeter, a search for high $p_T $ tracks and a matching between the clusters and the tracks. The selections applied at each trigger level are
as follows:
\begin{itemize}
\item{ {\bf Level 1}\\
The particles are selected using the Electromagnetic Calorimeter
information, applying cuts on the transvere energy ($E_T$) in the cluster and 
isolation criteria
around the clusters (using both hadronic and electromagnetic calorimeters, 
with reduced granularity information compared to that available in the HLT).
Each RoI is characterised by: $\eta, \phi, E_T ^{threshold}$ and isolation
criteria.

}
\item{ {\bf Level 2} \\
Starting from the LVL1 region of interest (RoI) of size
$\Delta\eta \times \Delta\phi = 0.2 \times 0.2$, the level 2 algorithms
refine the
selections using  the full granularity of the detector.
Electrons are selected using both the calorimeter information and 
tracking information. The shower shapes in the calorimeter and the
tracks reconstructed in the inner detector near the calorimeter clusters are
analyzed, applying selection cuts on $E/p$, $\Delta\eta$ and
$\Delta\phi$, as described in detail in table \ref{algo} 
(L2 Calo Hypo and L2 Track Hypo algorithms). 
}

\item{ {\bf Event Filter} \\
Starting from the Level 2 objects, the Event Filter refines the cuts using more
sophisticated algorithms, and access to  full event information in the 
calorimeter and the inner detector. The track search is performed in the SCT
and pixel detectors and independently in the TRT. Finally, the objects have
to pass three sets of cuts: EF Calo Hypo, EF Track Hypo and EF Match Hypo 
(checking the $E/p$ and spatial matching between the objects found in 
the tracking and calorimetry).
}
\end{itemize}

\begin{table}
\begin{center}
\begin{tabular}{|c|c|c|}
\hline
{\bf Hypothesis} & {\bf Cuts Applied }  & {\bf Cuts
  Applied } \\ 
{\bf Algorithm} & {\bf (e25i) }  & {\bf (2e15i) } \\ 
\hline
L2 Calo Hypo  & (E3x7/ E7x7) $>$ 0.88  & (E3x7/ E7x7)$>$ 0.9\\
 &(E1-E2)/ (E1+E2)$>$ 0.64 &(E1-E2)/ (E1+E2)$>$ 0.72\\

& ET (em)$>$ 22.0 GeV & ET (em)$>$ 11 GeV \\ 
& ET (had)$<$ 5.5 GeV & ET (had)$<$ 1 GeV \\ \hline

L2 Track Hypo&   PT$>$ 8 GeV &PT$>$ 8 GeV\\
 & 0$<ET/PT<$ 3.5&  0.2$<ET/PT<$ 3\\
& $\Delta\eta<0.08$ match & $\Delta\eta<0.07$ match \\ 
& $\Delta\phi<0.05$ match depending on $\eta$ & $\Delta\phi<0.04$ match depending on $\eta$   \\ \hline

EF Calo Hypo &  ET$>$ 23 GeV & ET$>$ 12.5 GeV\\
& $\Delta\eta$, $\Delta\phi <$ 0.099&  $\Delta\eta$, $\Delta\phi < 0.2$\\ \hline

EF Track Hypo & $N_{SCThits}>2$, $N_{blayerhits}>0$&${N_{SCThits}}>7$\\
& Impact.Par$<$ 0.5mm& Impact.Par$<$ 1mm \\ \hline

EF Match Hypo & 0.86$<ET/PT<$2.29, $\eta<1.37$ &0.7$<ET/PT<$ 1.7, $\eta <$1.37\\
&0.7$<ET/PT< $2.5, $\eta >$1.37 &0.7$<ET/PT<$ 2.5, $\eta >$1.37\\
&$\Delta\eta$ match$<$0.005, $\Delta\phi$ match$<$0.018& $\Delta\eta$ match$<$0.01, $\Delta\phi$ match$<$0.02 \\ \hline 
\end{tabular}
\end{center}
\caption{ \it Hypothesis Algorithms of the electron trigger chain with
the cuts applied at each step of the selection \cite{cibran_1}. \label{algo}}
\end{table}

\subsection{ Preselection Cuts for e25i and 2e15i: Definition of
  Reference Electrons \label{def}}
All the analyses have been performed using the trigger selection
hypothesis algorithms corresponding to the HLT trigger chain as described in
\ref{soft}, together with with various preselection criteria. These 
preselection criteria remove events which are not of interest, because they 
would not pass the level 1 trigger, would not pass offline reconstruction,
or because Monte Carlo truth information
shows they contain no electrons of interest.

At the end of each job, the TrigSteerMonitor prints a table with
the efficiency for each of the algorithms in the sequence. The
efficiencies are calculated with respect to reference electrons
(as defined in the job options). Typically the efficiency could be
calculated with respect to the Monte Carlo electrons, to the LVL1
preselection or to offline reconstructed electrons, either separately
or in combination. 

Moreover, as described in the section \ref{mc}, a filter selection
is applied at event generation level to remove events with no chance of
passing the level 1 trigger, and this has to be
taken into account when computing the total cross section.


\subsubsection{Monte Carlo Preselection Cuts \label{pre}}

For the signal samples (single electrons, W and Z decays), 
additional preselection cuts have been
applied on the electrons at the MonteCarlo truth level. These select only
electrons that have their momentun in a reasonable interval and that do
not cross the crack region. The requirements are:

\begin{itemize}
\item{ one generated electron in each RoI,}
\item{ two RoIs per event (only for the Z events),}
\item{ Monte Carlo truth $5 < P_T <100$ GeV}
\item{ $|\eta|<1.45$ and $|\eta|>1.55$,  $|\eta|<2.45$; these cuts exclude 
the crack region.}
\end{itemize}

\subsubsection{Offline Reconstructed Preselection Cuts \label{off}}

To compare the triggered electrons with the sample that would be 
reconstructed offline, the offline selection algorithms are also run
on all electron candidates. The first two algorithms 
that are run define as an offline electron any cluster-track match,
and then set a series of bits in the IsEM flag variable. For 
all electron candidates,
the candidate has to pass a series of cuts based on the shower shape
properties in different compartments of the calorimeter as well as
variables combining ID and calorimeter information. If a cut is not passed,
then a corresponding veto bit is set in the isEM flag. For candidates with an
associated track,
identification cuts based on the tracking information have to be
passed. Thus if isEM=0, then this is a good electron or photon.

\subsubsection{LVL1 Preselection Cuts \label{lvl1}}

The Level1 preselection algorithm simulates the decision of the LVL1 
trigger, applying these cuts \cite{lvl1}:
\begin{itemize}
\item{ClusterThreshold  = 19.0 GeV (e25i) or 9.0 GeV (2e15i)}
\item{EmRingIsolation = 3.0 GeV (e25i) or 8.0 GeV (2e15i)}
\item{HadRingIsolation = 2.0 GeV (e25i) or 4.0 GeV (2e15i)}
\item{HadCoreIsolation = 2.0 GeV (e25i) or 4.0 GeV (2e15i)}
\end{itemize}

\subsection{Datasets \label{datasets}}

Several different datasets were used to estimate the composition of the 
electron stream at low luminosity ($10^{33}\rm cm^{-2}s^{-1}$). 
We took samples generated and simulated for the ATLAS Physics Workshop, 
in particular: 
\begin{itemize}
\item{Single electrons with $E_T=25$ GeV, generated with Pythia, about 1000
  events\\ 
  ({\tt dataset rome.004022.xxx\_Pt\_25} Rome Production)}
\item{$Z \rightarrow e^{+} e^{-}$ generated with Pythia, about 10000
  events\\ 
  ({\tt dataset rome.004201.recolum01.ZeeJimmy} Rome Production)}
\item{$W \rightarrow \nu_{e} e^{-}$ generated with Pythia, about 10000
  events\\ 
  ({\tt dataset rome.004203.recolum01.WenuJimmy} Rome Production)}
\item{QCD di-jets generated with Pythia, about 138k events. This
  allows the evaluation of the trigger background \\
({\tt dataset rome.004814.recolum01.JF17\_pythia\_jet\_filter} Rome Production) }.
\end{itemize}

\section{Results}

In this section the results after the trigger selection are shown.
For each sample, the efficiency is calculated at every step
of the trigger chain. Both e25i and 2e15i are used.
The trigger chains have been run on the ESD samples used previously for the
Rome Physics Workshop, the datasets defined in the section \ref{datasets}.

The following efficiencies are defined with respect to both offline
reconstructed electrons and LVL1 confirmation, as described in detail
in section \ref{def}. 
The estimation of the rates have been performed for low luminosity,
using a MonteCarlo analysis on the CBNT rootples.

\subsection{Preselection at Generation Level: Cross Section Used \label{mc}}
An estimation of the expected trigger rate and composition of stream is given for low luminosity.
We calculated the event rate starting from the cross section (as derived from
Pythia) and taking into account the geometrical acceptance of the detector, 
considering only events that can be reconstructed.
In detail, the cross section of production at LHC of the events analyzed are 
reported in table \ref{cs}, with the filter efficiency:
\begin{itemize}
\item{ Electrons should have $p_T$ greater than 15 GeV and $|\eta|<3.0$}
\item{ Two such leptons are required for Z events, and one for W events}
\end{itemize}

To calculate the background trigger rates for electrons,
 we are using the Rome dataset 4814 for dijets with ET(hard) = 15 GeV and a 
particle level filter cut of $E_T = 17$ GeV. Each dataset contains QCD di-jets as well as physics processes like $W \rightarrow \nu_{e} e^{-}$, $Z \rightarrow e^{+} e^{-}$ and direct photon production,
which have been added to the QCD jets according to their cross
sections. The total cross section is reported in table \ref{cs}.
\begin{table}
\begin{center}
\begin{tabular}{|c|c|c|c|}
\hline
{\bf Event} & {\bf cross section } &  {\bf Filter $\epsilon$ }&  {\bf
  cross section after filter selection }  \\ \hline
 $Z \rightarrow e^{+} e^{-}$& 1603.8 pb & 61\% &978.3  pb \\ \hline
$W \rightarrow \nu_{e} e^{-}$& 17907 pb & 47.9\% & 8577.5 pb \\ \hline
QCD di-jets & 2.3 mb  & 14.3 \% & 0.16 mb \\ \hline

\end{tabular}
\end{center}
\caption{ \it The cross-sections for the processes used in this study, together
with the acceptance of the particle level filter (geometrical and Monte Carlo truth cuts) and the final cross-sections used for rate calculations.
\label{cs}}
\end{table}

The rate is calculated using:
\begin{equation}
 Rate = \frac{\sigma}{Filter}\frac{N_{sel}}{N_{all}}* L 
\label{rate}
\end{equation}
where the filter takes into account the particle level filter applied
at the generation level and reported in table \ref{cs}, while
$N_{sel}$ and $N_{all}$ are the number of events
selected by all the trigger chain and the total number of events respectively.

\subsection{Rates for Low Luminosity}
The rates below have been computed using the TrigESDAnalysis package
to estimate the efficiencies and purities after each trigger
algorithm (Athena Release 10.0.1, TrigESDAnalysis-00-00-05 tag), and
the MonteCarlo analysis using the CBNT Ntuples.  
Equation \ref{rate} has been used, with $L=\rm 10^{33}cm^{2}s^{-1}$ and the
cross sections after the filter shown in table \ref{cs}.

\subsubsection{ Single Electron}

For reference, we have estimated the trigger efficiency for a sample of single
electrons with $E_T=25$ GeV (e25i chain), with pile-up added.

The efficiency of preselection (including both LVL1 confirmation and
offline reconstruction isEM) for this sample is about 84\%.
Table \ref{ele} reports the efficiencies of each algorithm
with respect the previous one. These efficiencies are not cumulative.

\begin{table}
\begin{center}
\begin{tabular}{|c|c|c|}
\hline
{\bf Algorithm} & {\bf Number of Events Single e } &  {\bf Efficiency
  Single Algo. in {\%} } \\ \hline
 
Initial  & 932 & 100 \\ \hline 
ISEM PRES. & 790 & 84 \\ \hline
LVL1 PRES. & 784 & 99 \\ \hline
CALO LVL2 & 772 & 98 \\ \hline
ID LVL2& 736 & 95  \\ \hline
CALO EF &  706 & 95\\ \hline
ID EF & 698 & 98\\ \hline
Matching EF & 676 & 96 \\ \hline 
Cumulative Eff & 676/784 & 84 \\ \hline
\end{tabular}
\end{center}
\caption{ \it The values of efficiencies with respect to the
  reconstructed events and LVL1 confirmation for each algorithm
  separately. The same trigger chain has been used: e25i. The total
  number of events are about 1000 single electrons of 25 GeV. The
  number of events after each preselection algorithm are reported too.\label{ele}}
\end{table}

\begin{figure}[!h]

\begin{center}
\begin{tabular}{cc}
\epsfig{figure=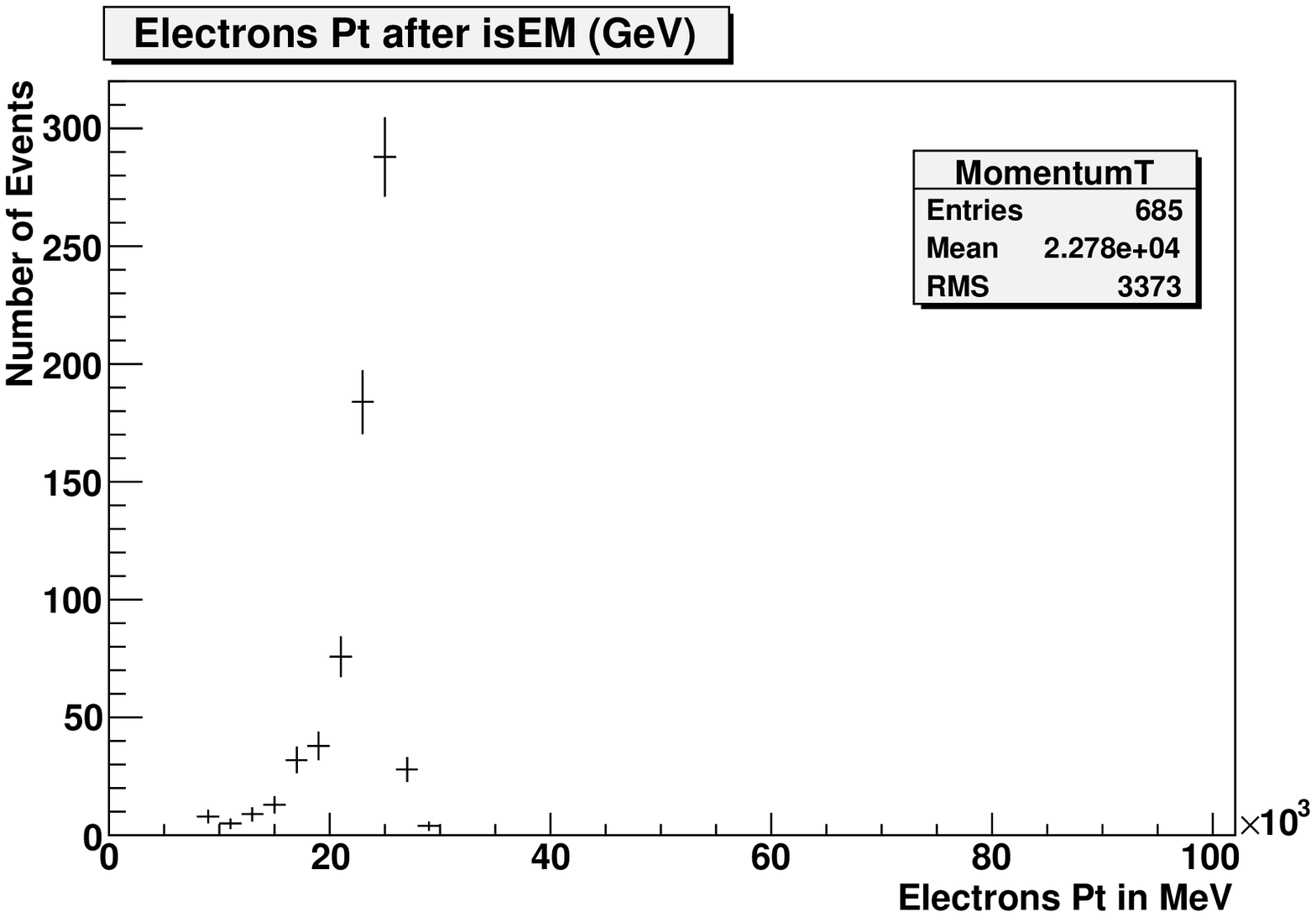,height=5.5cm,width=5.5cm} &
\epsfig{figure=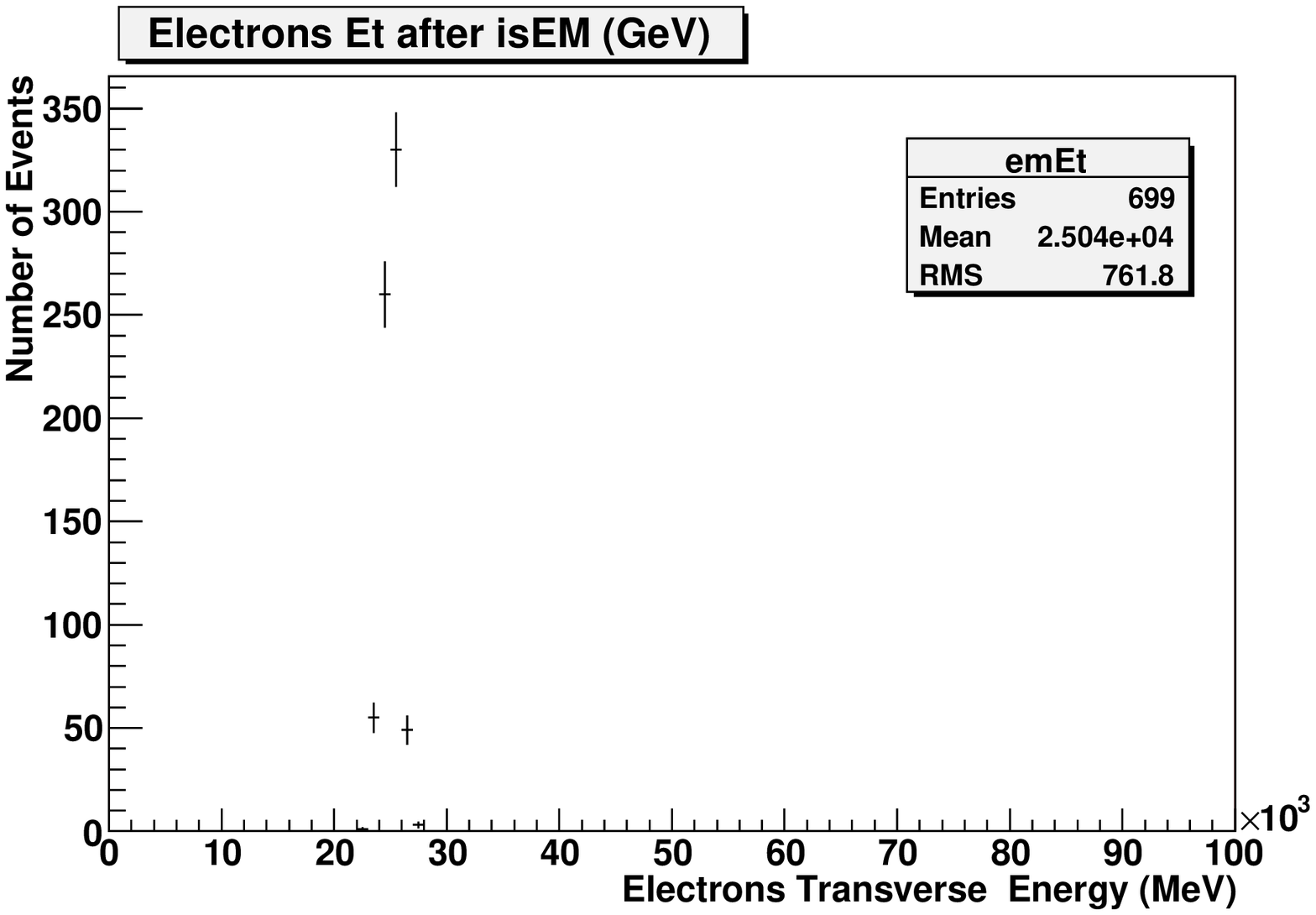 ,height=5.5cm,width=5.5cm} \\
\end{tabular}
\end{center}
\caption{\it Electron $p_T$ measured in the tracking (left) and 
  energy ($E_T$) in the calorimeter (right), after the isEM preselection 
cuts applied, for the $E_T=25$ GeV single electron sample. \label{single}}
\end{figure}

Figure \ref{single} shows the $p_T$ and energy
distributions after the preselection cut isEM.

\subsubsection{$W \rightarrow e \nu_e$}
The sample W is selected applying the trigger chain: e25i.
The electrons in the crack region have been excluded and only one
electron in each electromagnetic cluster is required too (Number of
primary electrons equal to one). The isEM flag and the LVL1
confirmation are then applied as preselection to define the right sample of
reference electrons.
Table \ref{w} shows the results for the W sample, compared with the
efficiencies derived from the 25 GeV single electron sample.
The efficiencies are cumulative, and calculated with respect to the
electrons obtained after the preselection cuts. 
                                           
The estimated rate after all selections, using the formula \ref{rate} 
and the cross section reported in table \ref{cs}, is about 9.1 Hz at 
a luminosity of $L=\rm 10^{33}cm^{-2}s^{-1}$.

\begin{figure}[!h]
\begin{center}
\epsfig{figure=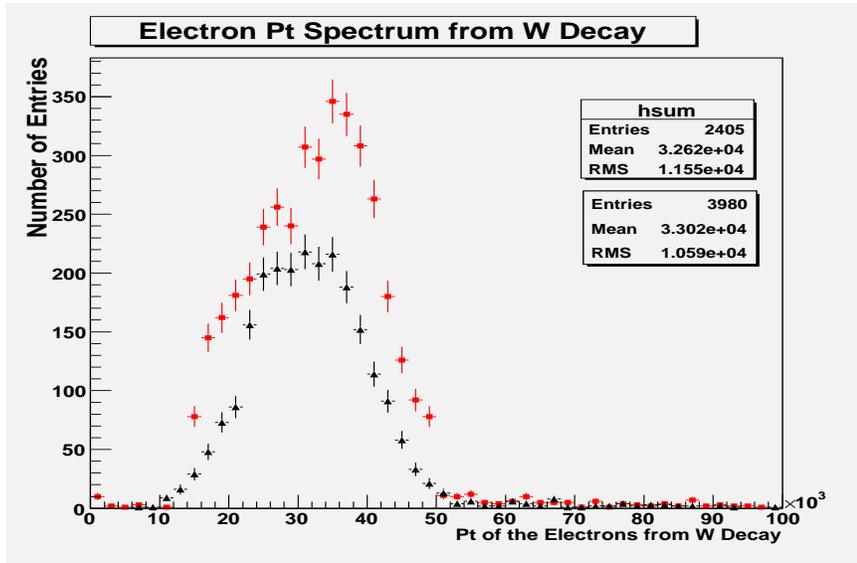,height=7.5cm,width=11.5cm} 
\end{center}
\caption{\it  
  Electrons $P_T$ spectrum of events $W
  \rightarrow \nu_{e} e^{-}$. In black triangles there is the electrons
  $P_T$ spectrum of events after all the trigger chain e25i
  reconstructed in the tracker, in red squares
the distribution obtained after
  the preselection cuts and before the selection of the e25i chain,
  these are the ``reference events''. 
   \label{w1}    }
\end{figure}

In figure \ref{w1}, the $p_T$ distributions of the
electrons from W decay after the preselection and the complete e25i trigger
chain are shown.

\begin{figure}[!h]
\begin{center}
\epsfig{figure=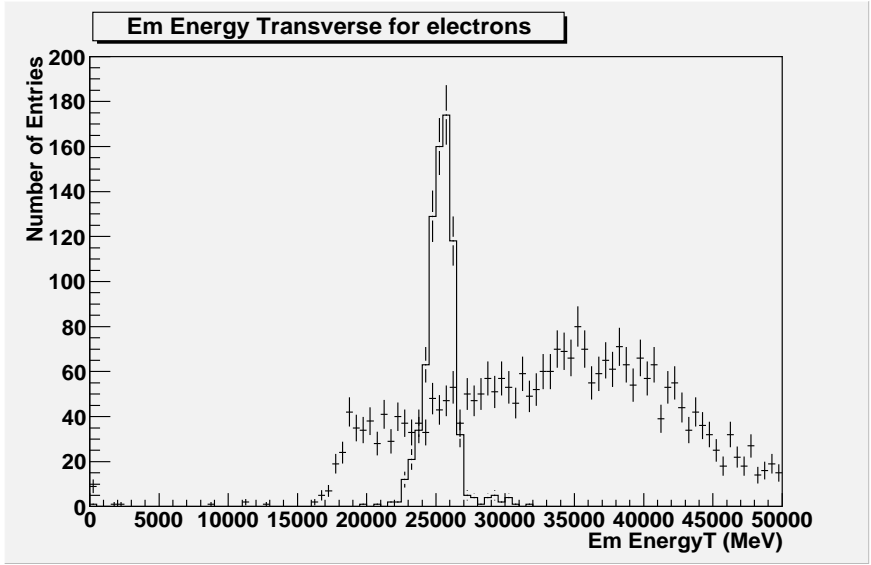,height=7.5cm,width=11.5cm} 
\end{center}
\caption{\it  
 Transverse electromagnetic energy spectrum from $W
  \rightarrow \nu_{e} e^{-}$ events (points) and single electrons (histo).
The threshold defined in the e25i trigger chain is $EmE_T > 22 GeV$,
  this explains the different algorithm efficiency between
  single electrons and W events.  
   \label{eme}    }
\end{figure}

The comparison between the electromagnetic energy transverse for the
single electrons and the electrons from W decay is shown in the plot
\ref{eme}, this explains the different efficiency obtained for the
T2CaloHypo algorithm for these events. 
The efficiency as a function of $p_T$ of the electrons
from W decay is shown in figure \ref{w2}.  
 
The $\eta$ distributions for single electrons after the Monte Carlo preselections  
are shown in figure \ref{w2}.

\begin{figure}[!h]
\begin{center}
\epsfig{figure=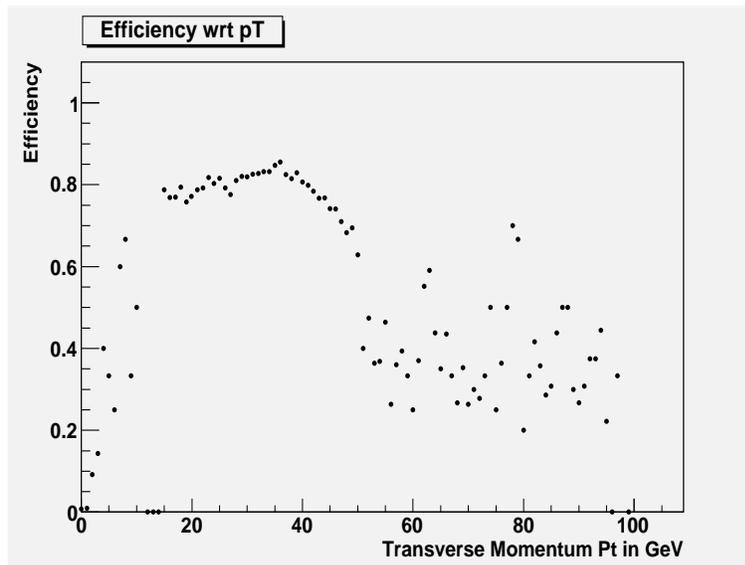,height=7.5cm,width=10cm} 
\end{center}
\caption{\it  
Efficiency with respect to the transverse momentum of the   
electrons from events $W
  \rightarrow \nu_{e} e^{-}$. 
The efficiency is calculated per bin (1GeV), as the ratio of the
MonteCarlo $p_T$ spectrum before and after all the cuts.
   \label{w2}    }
\end{figure}

\begin{figure}[!h]

\begin{center}
\begin{tabular}{cc}
\epsfig{figure=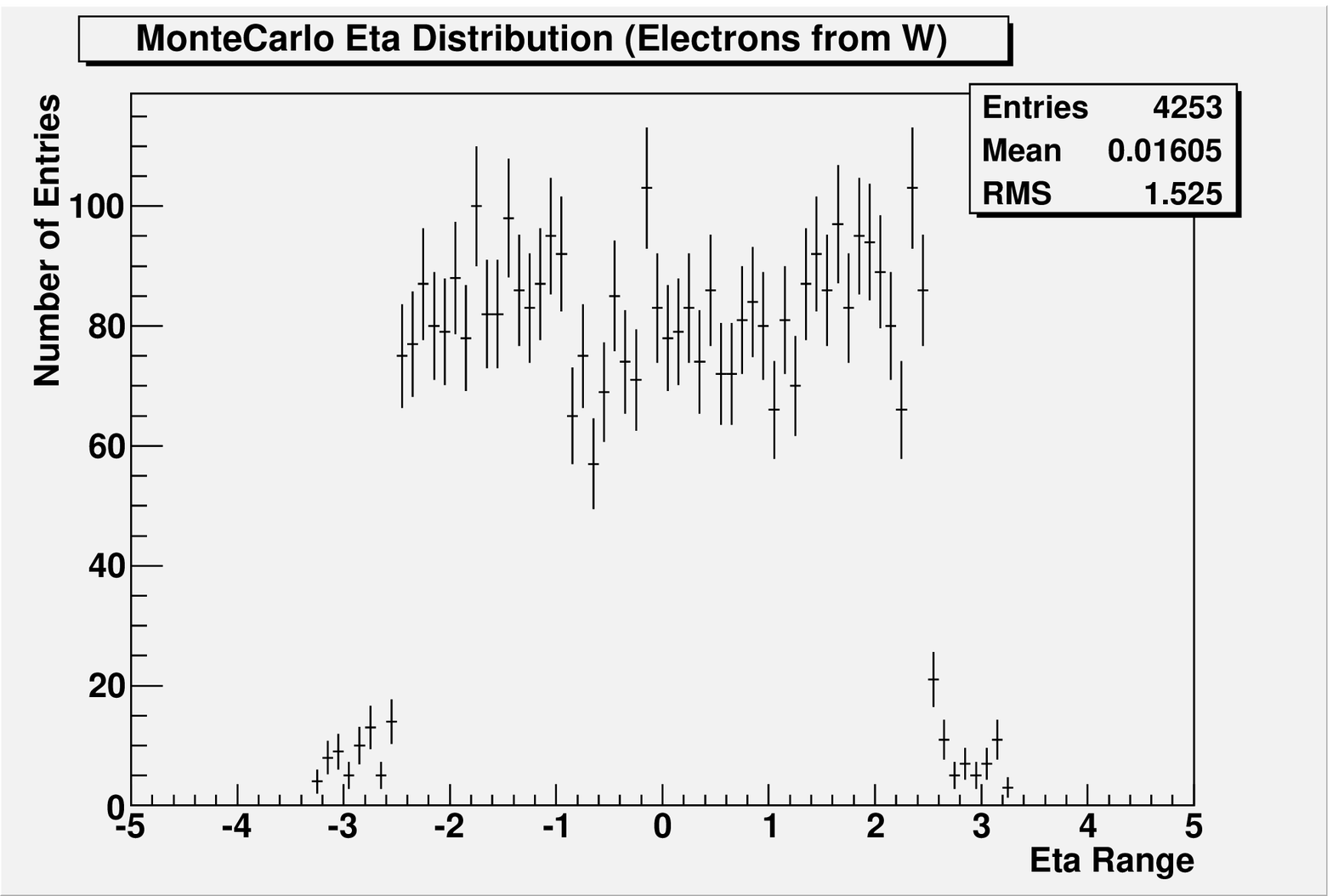,height=5.5cm,width=5.5cm} &
\epsfig{figure=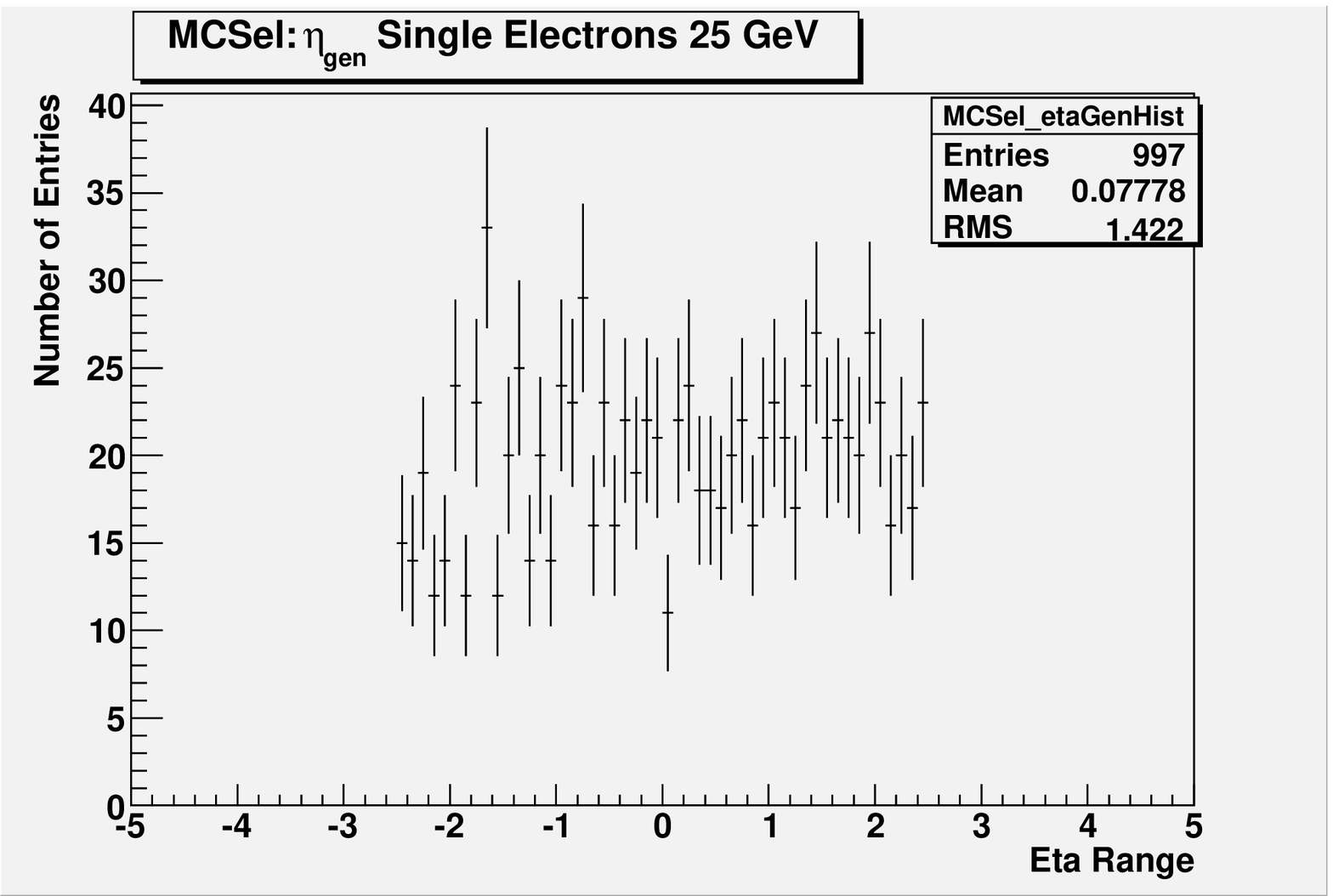,height=5.5cm,width=5.5cm} \\
(a)& (b)\\
\end{tabular}
\end{center}
\caption{\it The (a) plot shows the
the electrons $\eta$ distribution from MonteCarlo
informations (with the constraint of RoI) for the sample: $W
\rightarrow \nu_{e} e^{-}$ while the (b) plot the same distribution
but for the single electrons of 25 GeV. \label{w2}}
\end{figure}

\begin{table}
\begin{center}
\begin{tabular}{|c|c|c|c|}
\hline
{\bf Algorithm} & {\bf Number of W events }& {\bf Efficiency W in {\%} } &  {\bf Efficiency
  Single e in {\%} } \\ \hline
 
Initial & 10036 &100 & 100 \\ \hline 
ISEM PRES. & 5329 &53 & 84 \\ \hline
LVL1 PRES. & 5101 & 51 & 84 \\ \hline
CALO LVL2 & 4470& 88 & 98 \\ \hline
ID LVL2& 4156& 81 & 94  \\ \hline
CALO EF &  3944& 77 & 90\\ \hline
ID EF & 3904& 76 & 89\\ \hline
Matching EF & 3768&74  & 86  \\ \hline 
\end{tabular}
\end{center}
\caption{ \it The values of efficiencies with respect to the
  reconstructed events (after LVL preselection) for all the algorithms (ID and Calo) for the
  level 2 and the event filter. The cumulative
  efficiencies of $W \rightarrow \nu_{e} e^{-}$ and
  $E_T=25$ GeV single electron samples are compared, for the e25i trigger
chain.  \label{w}}
\end{table}

\newpage

\subsubsection{$Z \rightarrow e^+ e^- $}
The Z sample 
is selected applying two independent trigger chains: e25i and 2e15i.
In both cases the electrons in the crack region have been excluded and
only one electron in each electromagnetic cluster is required
(Number of primary electrons equal to one). 
The Monte Carlo preselection cuts have been applied before the trigger
chains, see section \ref{pre}.
The trigger steering uses two independent selection chains, and the number
of selected events is the algebraic union of the events selected by
the two chains. 
In the table \ref{z}, all the efficiencies are reported for each
algorithm, and finally the total efficiency of selection obtained by
the sum of the e25i and 2e15i trigger chains. 
For each algorithm is reported the number of selected events, an event
is defined ``selected'' with respect to the 2e15i chain when there are two
electrons per event while for the e25i chain when there are one or two
electrons per event. 

The estimated rate using the formula \ref{rate} and the cross section
after filter reported in table \ref{cs}, is about 0.84 Hz. 
The combined efficiency is obtained taking the events that are
selected by e25i or 2e15i trigger chain and excluding double counting from
events selected by both. It is interesting to note that nearly all Z events
are selected by the e25i trigger chain alone. The small fraction of electron
momenta betwen 15 and 25 GeV and the lower efficiency for selecting two
electrons as compared to one mean that the 2e15i trigger adds only a small 
number of events not selected by e25i, and 2e15i alone has a significantly
lower efficiency for the overall sample.

\begin{table}
\begin{center}
\begin{tabular}{|c|c|c|}
\hline
{\bf Algorithm} & {\bf Number of Events } & {\bf Number of Events }  \\ \hline
 & {\bf e25i }   & {\bf 2e15i } \\ \hline
PRES. & 5694 & 5659 \\ \hline
CALO LVL2 & 5543& 4969 \\ \hline
ID LVL2& 5304 & 4969  \\ \hline
CALO EF & 5160 &4940\\ \hline
ID EF & 4993& 4288\\ \hline
Matching EF &4955 & 4003 \\ \hline 
Efficiency &87.6\% &70.3\% \\\hline
Combined Eff. & \multicolumn{2}{c|}{88\%} \\\hline



\end{tabular}
\end{center}
\caption{ \it The number of   $\rm Z\rightarrow e^{+} e^{-}$ 
events for each single algorithm of the
  trigger chains: 2e15i and e25i, and their matching. The last rows show
the efficiency for each trigger chain and their
  combination, the number of events of the match is calculated summing
  the events accepted by a trigger chain plus the events of the other
  chain not yet included.   
Almost all of the events are triggered by e25i trigger chain.
\label{z}    }
\end{table}

\begin{figure}[!h]

\begin{center}
\begin{tabular}{cc}
\epsfig{figure=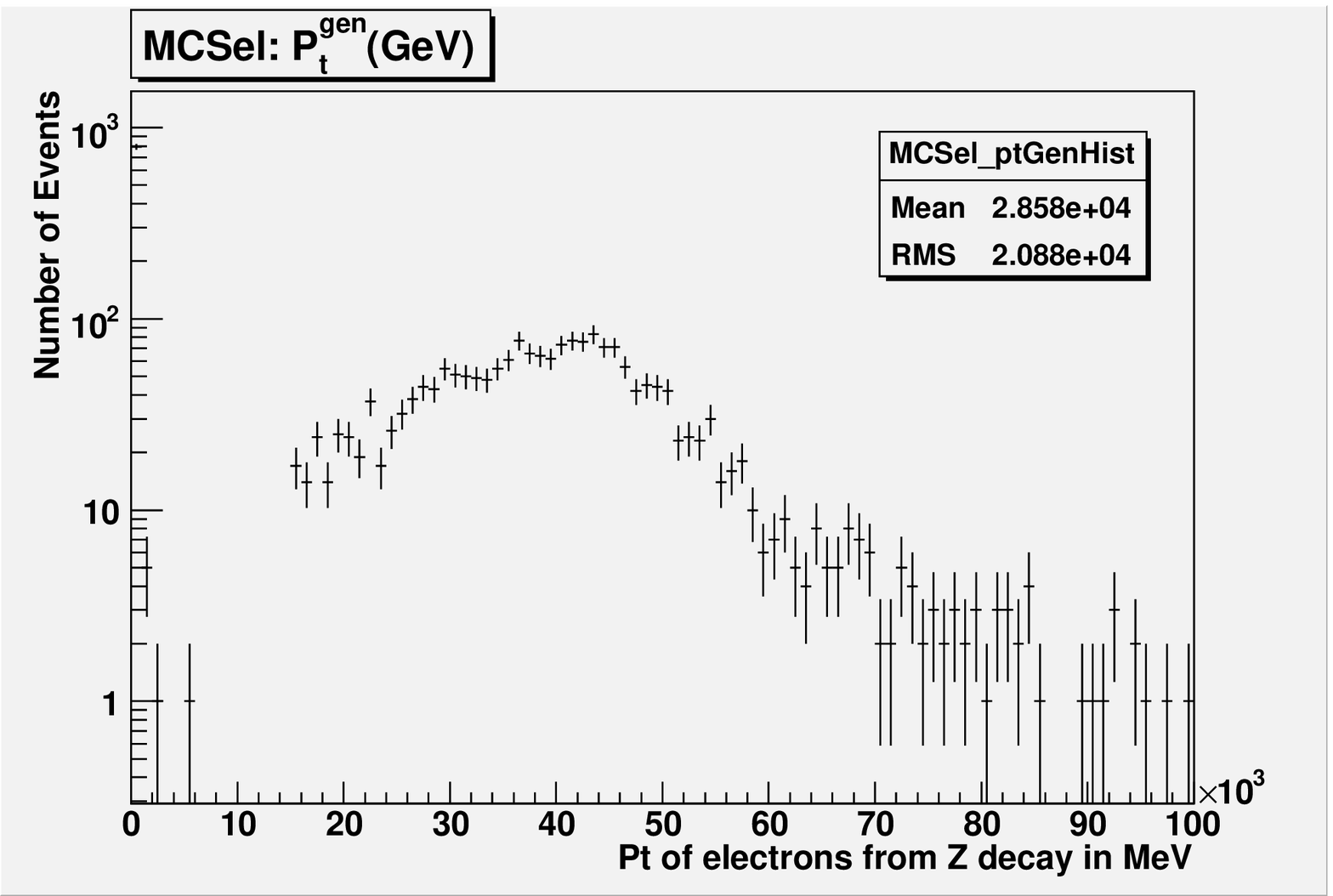,height=5.5cm,width=5.5cm} &
\epsfig{figure=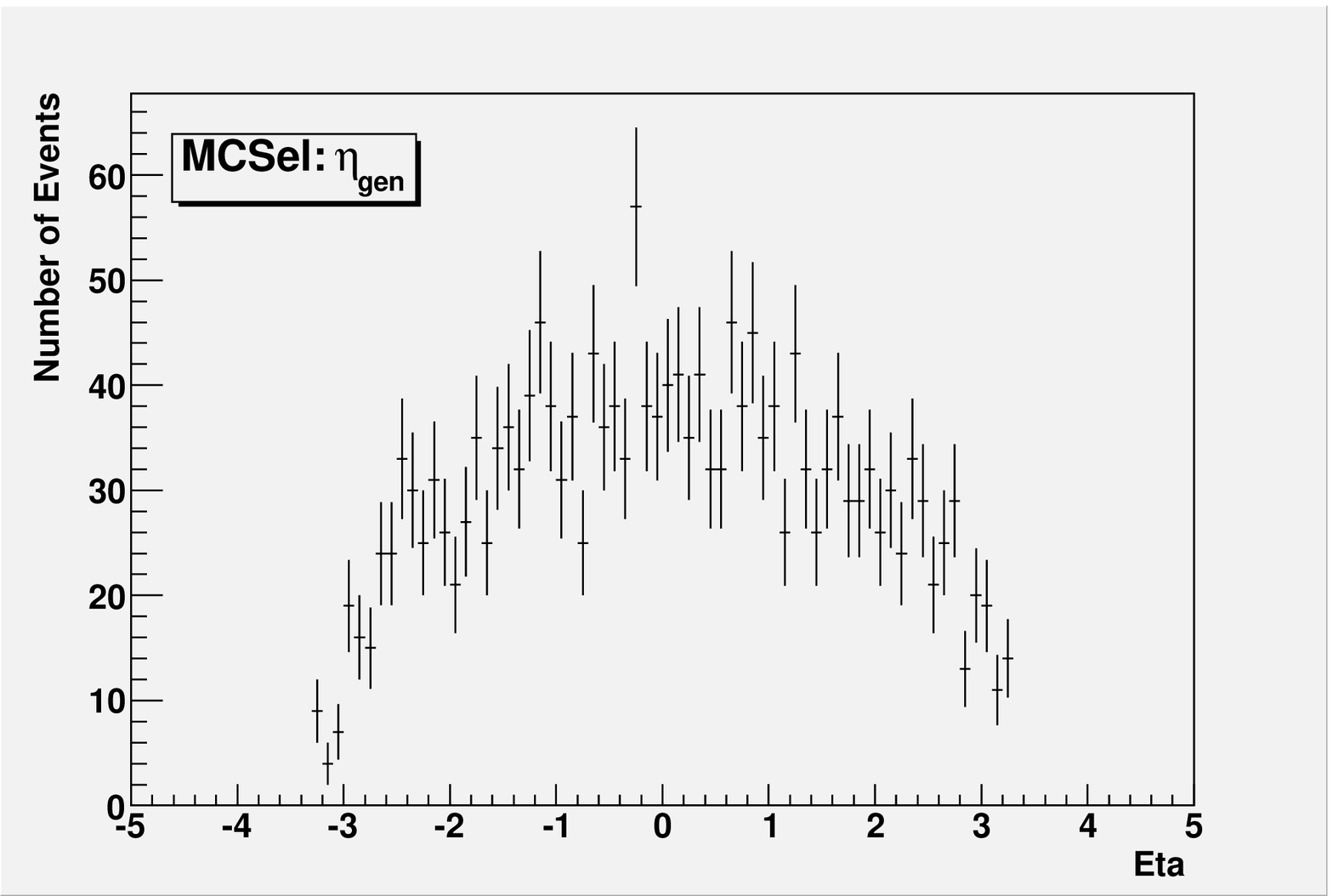,height=5.5cm,width=5.5cm} \\
(a)& (b)\\
\epsfig{figure=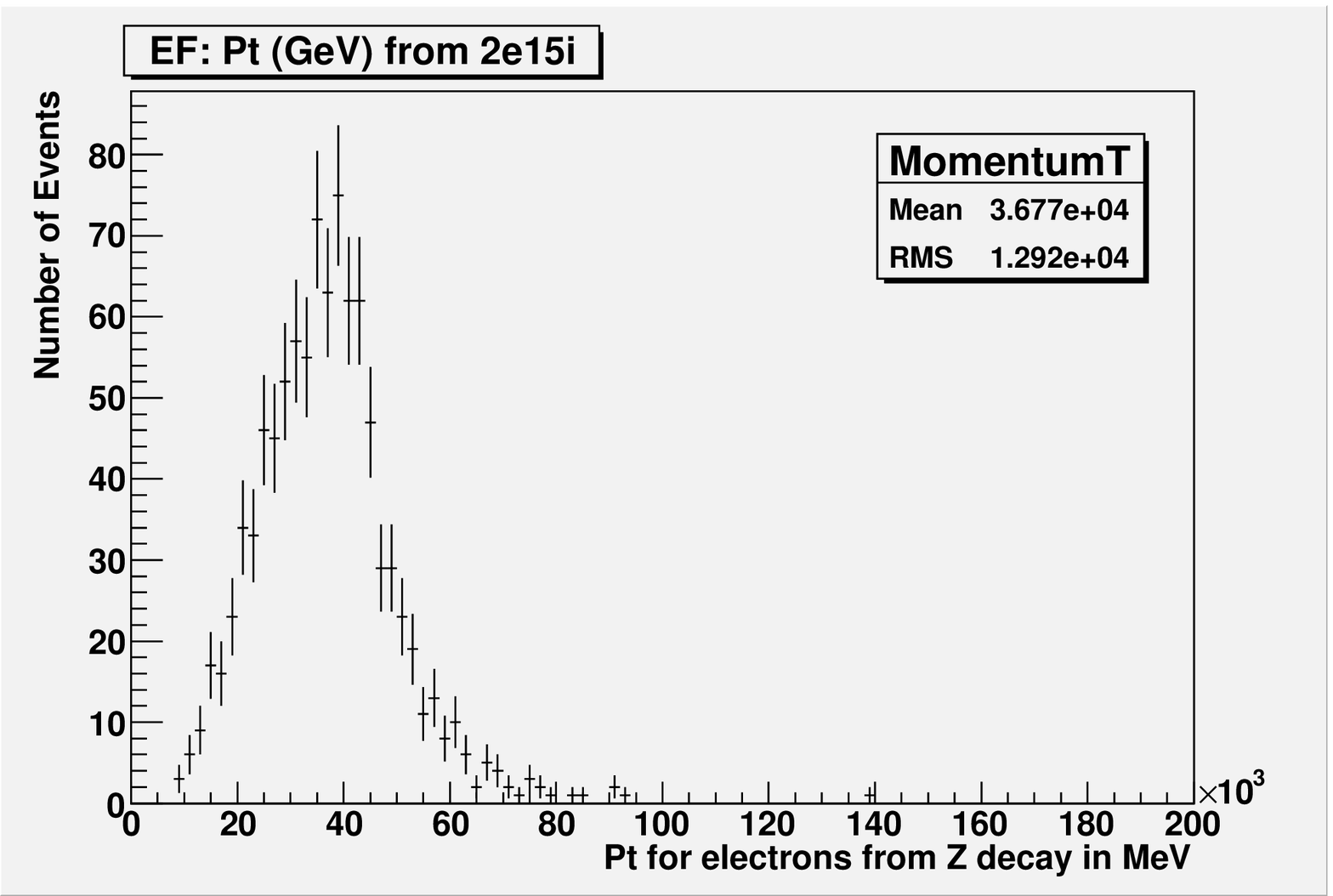,height=5.5cm,width=5.5cm} &
\epsfig{figure=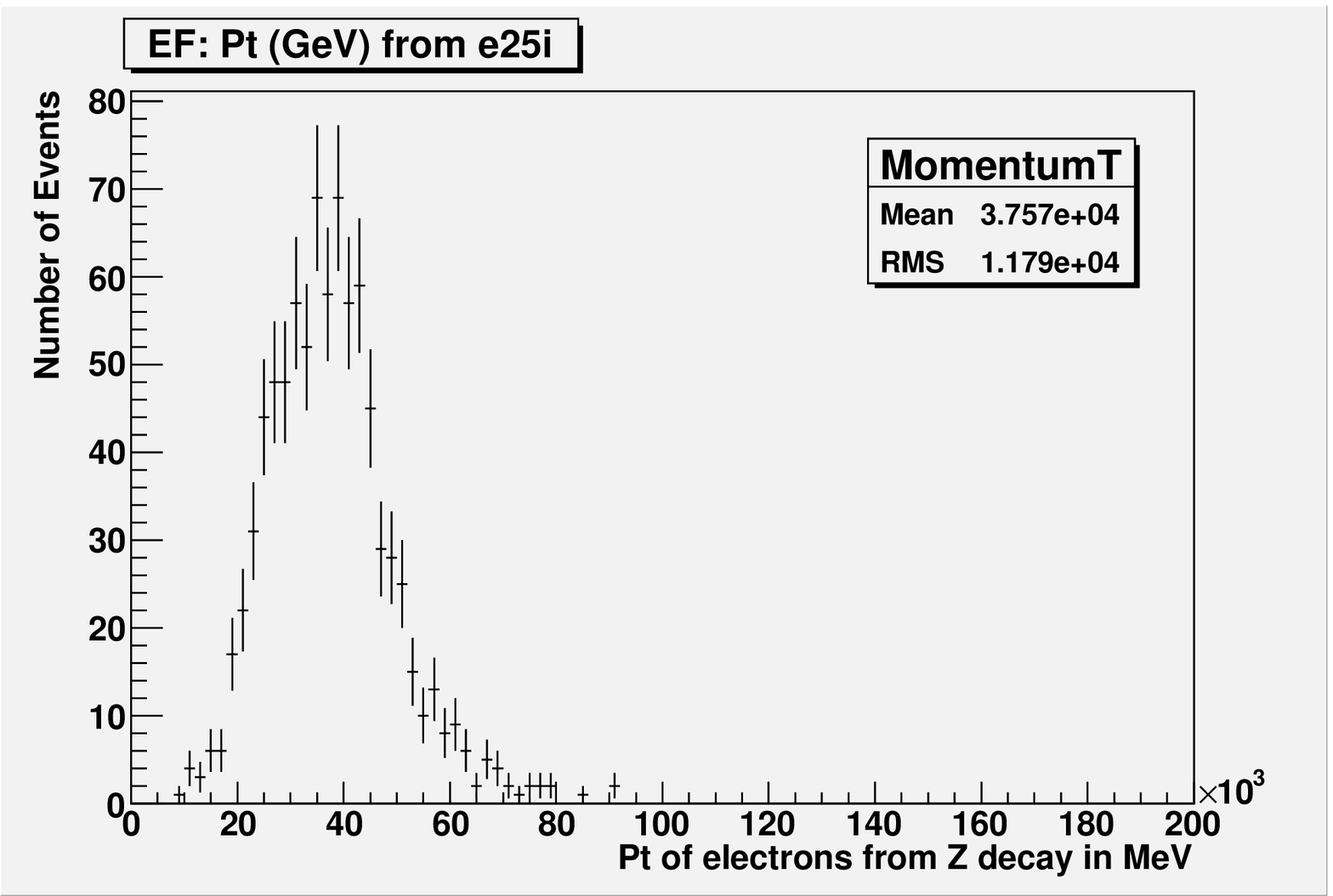,height=5.5cm,width=5.5cm} \\
(c)& (d)\\
\end{tabular}
\end{center}
\caption{\it  $\rm Z\rightarrow e^{+} e^{-}$ sample. The (a) plot shows the electron $p_T$ from MonteCarlo
  information (with the constraint of RoI), while the (b) plot
  shows the eta distribution of the electrons from Z decay.
  The electron $p_T$ distributions, after the trigger chains: 2e15i
  and e25i, are reported in the plots (c) and (d) respectively.\label{zeta}}
\end{figure}


\subsubsection{QCD Di-jets}
The selection of QCD events is done using the same hypothesis algorithms 
as for the signal samples, but without the requirement on the number of primary electrons in each cluster.
The electron candidates in this sample represent the background in the calibration stream with respect the other physics events described above.
Each dataset contains QCD di-jets as well as physics processes like $W \rightarrow \nu_{e} e^{-}$, $Z \rightarrow e^{+} e^{-}$, and direct photon 
production, so we analyzed in detail the Monte Carlo composition of the stream to define what we obtain after the selection of the relevant physics processes.
Applying the cuts described above the expected rate is about 20 Hz, with an efficiency of 2.9 \% with respect to the preselection and an efficiency of preselection of about 0.004 \%.

The total number of analysed events total events is about 140000, and only
16 events survive all the trigger cuts, with the following composition:
\begin{itemize}  
\item{Genuine electrons from W and Z or B hadron decays about 50\%}
\item{Converted photons from $\pi$ or jets about 31\%}
\item{Fake events, for example charge particles with tracks randomly
associated to electromagnetic calorimeter clusters, about 19\%}
\end{itemize}

The selected events in the background sample that contain genuine electrons can
be considered as useful for the electron calibration stream as well. 

\begin{figure}[!h]
\begin{center}
\epsfig{figure=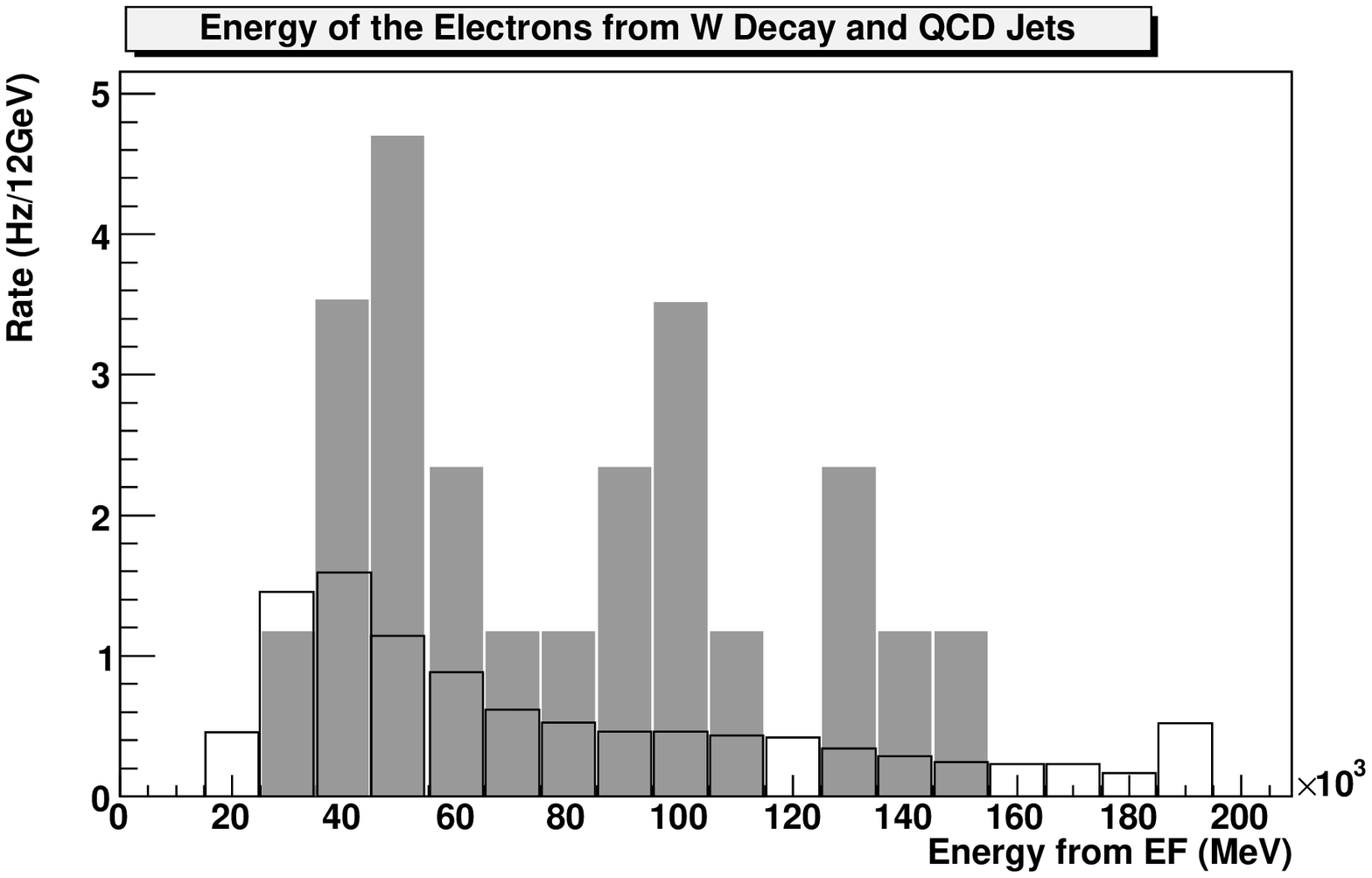,height=7.5cm,width=8.5cm} 
\caption{\it The plot shows the rates as function of electron energy
  after EF.
The rates have been obtained normalized the number of the entries with
respect to the rate with 12 GeV per bin. The QCD events are the
dark histogram, the electrons are represented by the open histogram.\label{energyrate}}
\end{center}
\end{figure}

Figure~\ref{energyrate} shows the rate of accepted events as a function of the 
energy of the triggered electron candidate from the Calorimeter EF algorithm. Although 
the overall statistics are very low, it can be seen that the electrons
from W and Z boson decays have a somewhat softer $E_T$ spectrum than those
from the QCD background.



\begin{table}
\begin{center}
\begin{tabular}{|c|c|c|}
\hline
{\bf Algorithm (e25i)} & {\bf Number of Events } & {\bf Rate (Hz) }  \\ \hline
Initial & 138532 & --- \\\hline
PRES. & 551 & 640  \\ \hline
CALO LVL2 & 170& 198 \\ \hline
ID LVL2& 33 & 38  \\ \hline
CALO EF & 25 & 29\\ \hline
ID EF & 25& 29\\ \hline
Matching EF &16 & 19 \\ \hline 

\end{tabular}
\end{center}
\caption{ \it The number of events for each single algorithm of the
  trigger chain 25ei, and their matching for QCD-jets events. \label{jet}}
\end{table}

\section{Different Selections}

\begin{figure}[!h]

\begin{center}
\begin{tabular}{cc}
\epsfig{figure=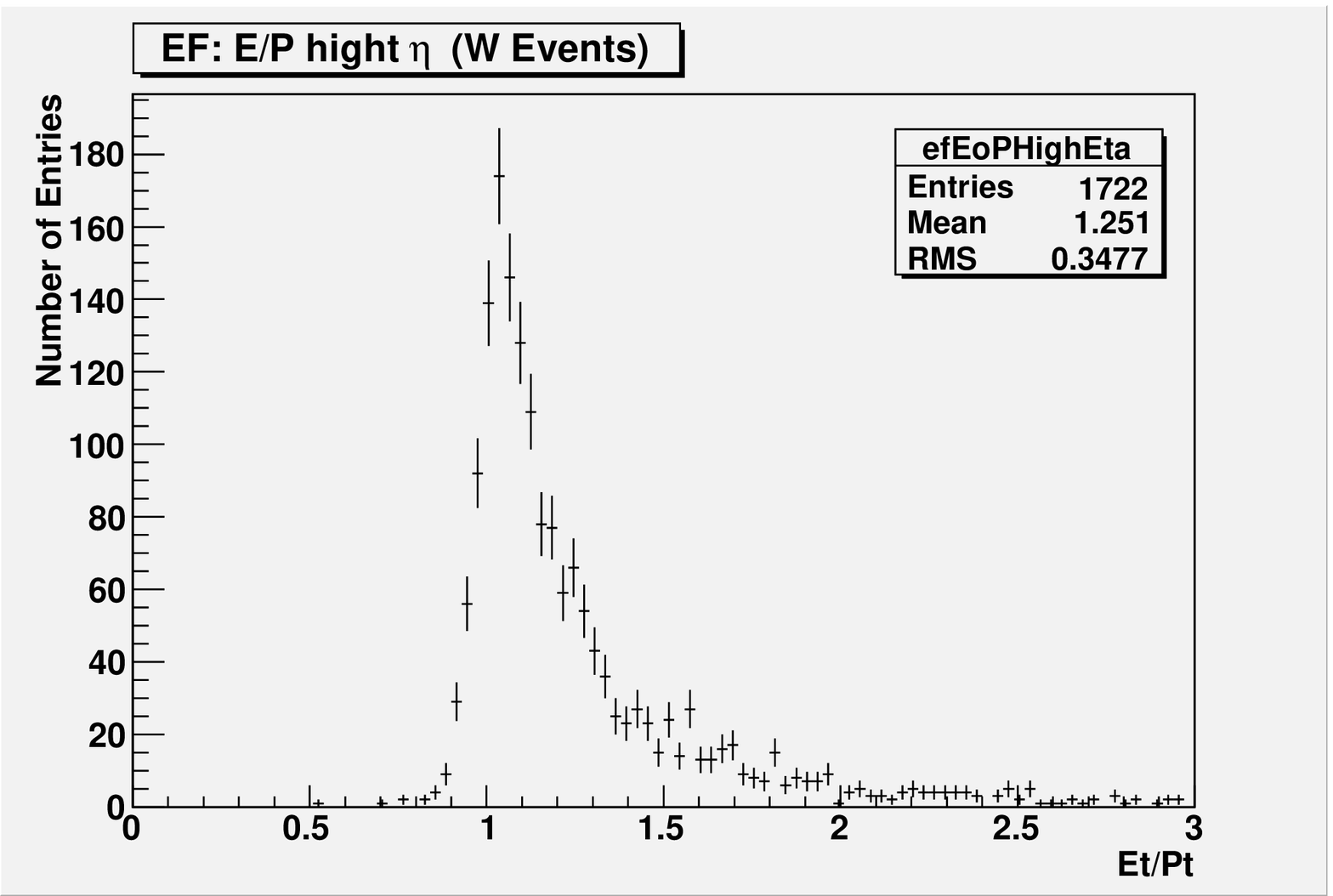,height=5.5cm,width=5.5cm} &
\epsfig{figure=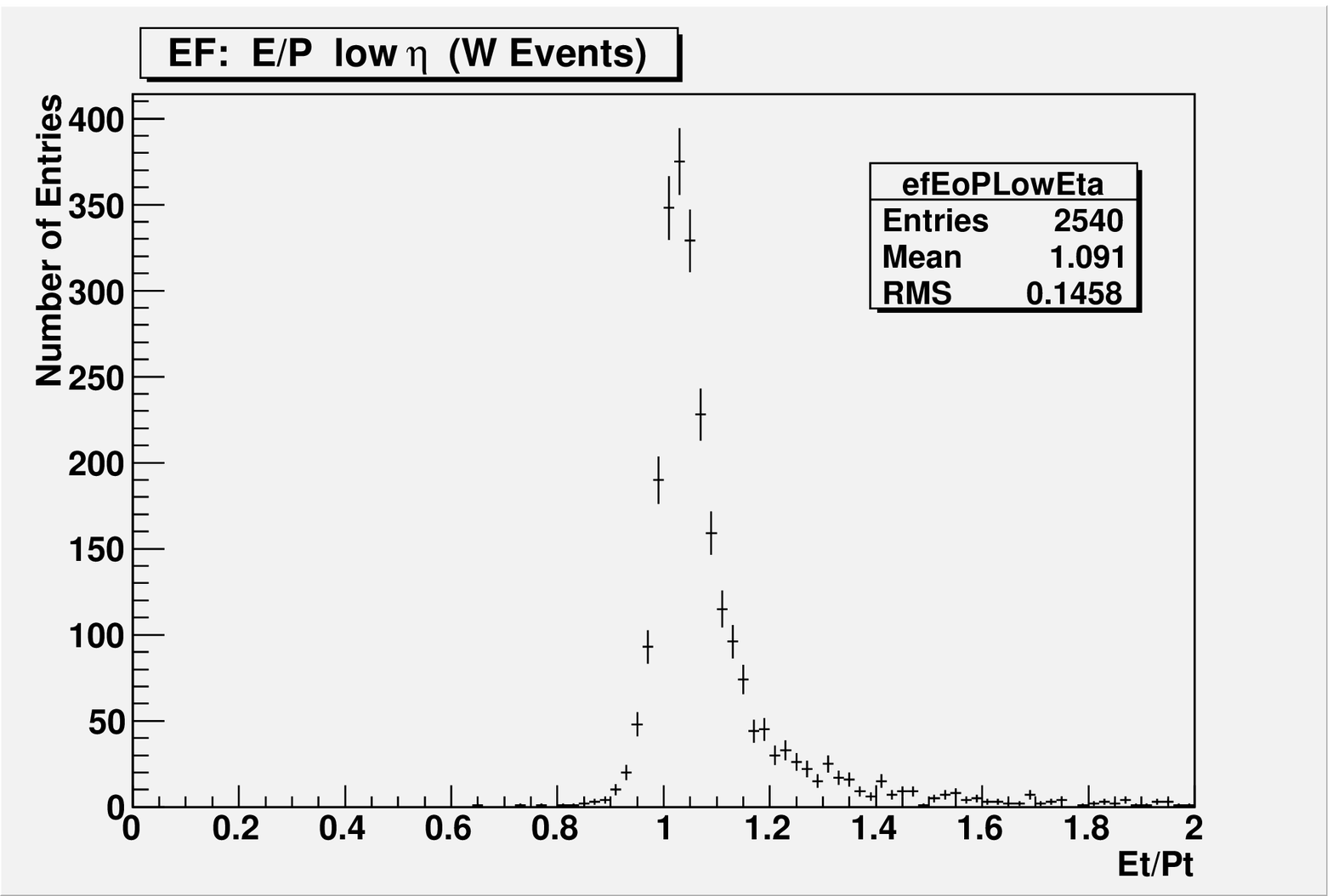,height=5.5cm,width=5.5cm} \\
(a)& (b) \\
\epsfig{figure=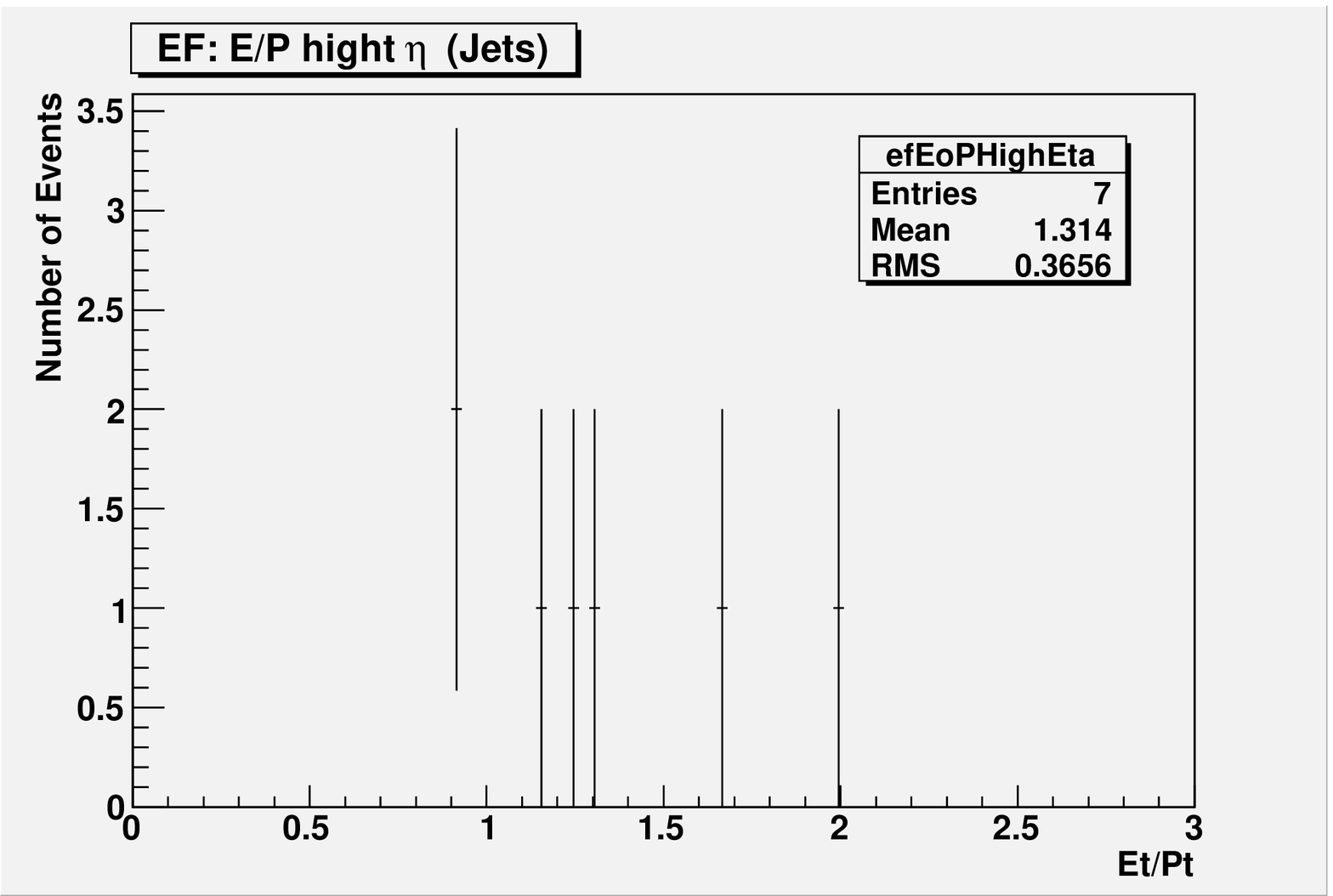,height=5.5cm,width=5.5cm} &
\epsfig{figure=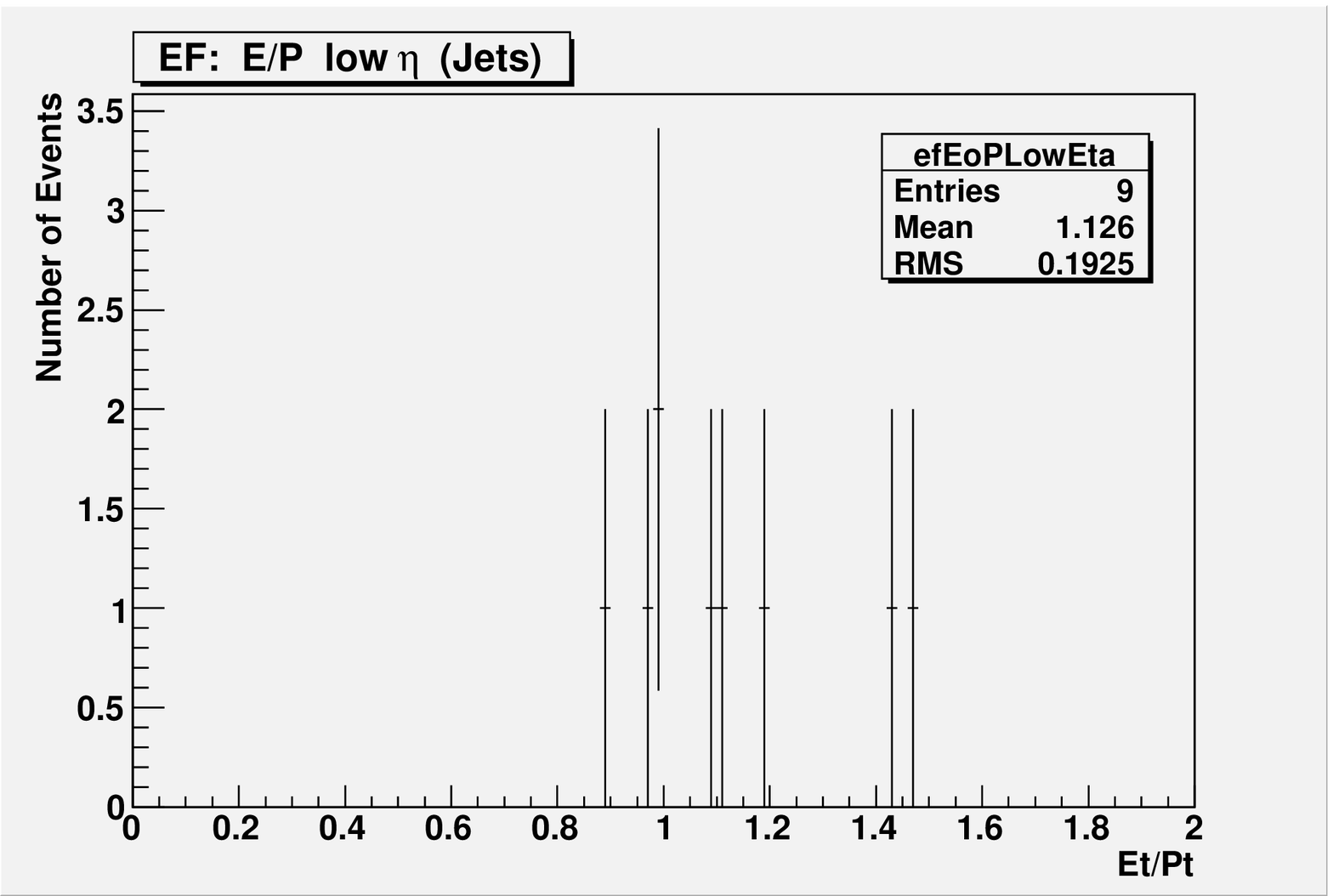,height=5.5cm,width=5.5cm} \\
(c)& (d) \\
\end{tabular}
\end{center}
\caption{\it In the (a) plot, the E/p ratio for the W events for high
  $\eta$ is
  shown, while in the (b) side the same E/p ratio for W but for the
  low $\eta$ cuts. In the (c) and (d) plots the same variables but for the
  electrons in QCD jets for high and low eta range. \label{ep}}
\end{figure}

In this section we studied applying different cuts at the event filter to
reduce the total electron stream trigger rate to 10 Hz, keeping as
large a fraction as possible of the pure electrons.  
The cuts applied at the calorimeter filter level are not yet optimized, and
possible improvements can be achieved modifying the thresholds
applied to the $E/p$ ratio and track parameters. Figure~\ref{ep} shows the
$E/p$ variable for W signal and QCD background events in different $\eta$ 
ranges. Again, although the statistics for the background sample are very 
low, it looks possible to increase the signal purity by tightening the
upper cut on $E/p$.

In table \ref{cuts}, the effect of applying two different sets of cuts is shown
on the rate of W and Z events in comparison with the rate of QCD jets.
Moreover for the QCD jets, the rate composition has been analysed and
the genuine electron component calculated.

\begin{table}
\begin{center}
\begin{tabular}{|c|c|c|c|c|}
\hline
{\bf EF Match Cuts} & {\bf W Event Rate } &  {\bf QCD Jets Rate }& {\bf Z Rate }& {\bf
  Total Rate }\\ \hline
0.86$<ET/PT<$2.29, $\eta>1.37$ &9.1Hz & 19Hz(9.5Hz) & 0.84Hz &28.9Hz(19.4Hz)\\
0.7$<ET/PT< $2.5, $\eta <$1.37 & & & &\\
\hline
0.86$<ET/PT<$1.65, $\eta> 1.37$ & 8.5 Hz & 13Hz(7.6Hz) & 0.88Hz & 22.4Hz(17.0Hz)\\ 
0.7$<ET/PT< $1.4, $\eta <$1.37 & & & &\\
\hline

0.86$<ET/PT< $1.65, $\eta >$1.37 & 7.3Hz & 9Hz (6.8Hz) & 0.87Hz & 17.2Hz(15.0Hz) \\
0.97$<ET/PT<$1.15, $\eta<1.37$ & & & &\\

\hline
\end{tabular}
\end{center}
\caption{ \it New cuts applied at Event Filter level to reduce the
  total rate to about 15 Hz. In the QCD column, in parenthesis, the
  rate of genuine electrons is reported. The Z rate is almost the
  same, the different cuts are applied only on the e25i chain then the
  matching of the two independent trigger chains reproduces the same rate.\label{cuts}}
\end{table}

Taking into account the last cuts, we can obtain about more than 17 Hz (Z+W+jet)
trigger rate, where about 15 Hz are genuine electrons useful for the
electron calibration stream. The Z events are less than 1 Hz after the
cuts.

\section{Conclusions and outlook}
After this first analysis on the Rome data sample using the Electron High Level
Trigger chains (e25i and 2e15i) we have estimated the electron trigger
rate expected at low luminosity.
The efficiency and the purity have been calculated, with the standard
thresholds and with modified cuts on the Event Filter, to optimize as
much as possible the ratio purity over efficiency of the obtained
sample. After some limited optimisation, a total rate of 17 Hz has been found, 
dominated by genuine electrons from W and Z decay.

For the background, the number of selected events is very small, due to the
limited size of the available background sample. This makes it impossible to
study further cut optimisation to reduce the rate to the target of 10 Hz. 
However, it is already apparent that simply increasing the $E_T$ threshold, 
while reducing the rate, will not increase the purity of the selected sample.
Much more background event statistics will be required to better
understand the relations between the background composition and the cuts on 
the $E/p$ ratio. 

More analysis will be performed on the CSC samples. In addition to looking
at the e25i trigger, it will be of interest to look at lower thresholds for 
lower luminosity running (e.g. $10^{31}$ and $10^{32} \rm cm^{-2}s^{-1}$). This
is not possible with the present background sample due to the filter cuts
imposed in the event generation.

\clearpage



\begin{thebibliography}{40}


\bibitem {nota1} R.Hawkings et F.Gianotti \emph{ATLAS detector calibration
  model: preliminary subdetector requirements}, ATL-GEN-INT-2005-001
\bibitem {hlt} ATLAS Collaboration \emph{ATLAS High-Level Triggers,
  DAQ and DCS Technical Proposal}, CERN/LHCC/2000-17, (2000)
\bibitem {lvl1} ATLAS Level1 Calo Group \emph{ATLAS Level-1
  Calorimeter Trigger Algorithms}, ATL-DAQ-2004-011, (2004)
\bibitem {atr_1} J.Baines et al. \emph{Perfomance Studies of the High
  Level Electron Trigger}, ATL-COM-DAQ-2003-020
\bibitem {atr_2} A.Gesualdi Mello et al. \emph{Overview of the
  High-Level Trigger Electron Photon Selection for the ATLAS
  Experiment at the LHC }, proccedings

\bibitem{cibran_1} \emph{Electron Trigger Optimization}, P Conde
  Muino, I. Grabowsky C. Padilla, E. Perez-Codina, C. Santamarina and
  G. Tetlalmatzi, e-gamma meeting, 26th of October 2005 

\end{thebibliography}
\end{document}